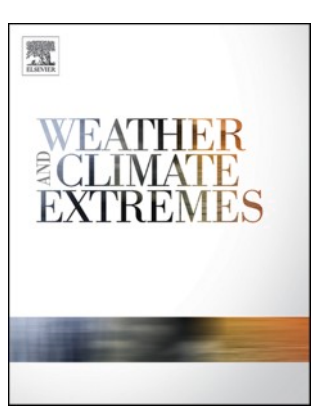

# Impact of extremely high temperature on future photovoltaic power potential over East Asia

Changyong Park[a], Ana Juzbašić[a], Dong-Hyun Cha[a,*], Seung-Ki Min[b,**], Joong-Bae Ahn[c], Eun-Chul Chang[d], Young-Hwa Byun[e], Youngeun Choi[f]

[a] Department of Civil, Urban, Earth, and Environmental Engineering, Ulsan National Institute of Science and Technology, 44919, Republic of Korea
[b] Division of Environmental Science and Engineering, Pohang University of Science and Technology, 37673, Republic of Korea
[c] Department of Atmospheric Sciences, Pusan National University, 46241, Republic of Korea
[d] Department of Atmospheric Science, Kongju National University, 32588, Republic of Korea
[e] Department of Atmospheric Sciences, Yonsei University, 03722, Republic of Korea
[f] Department of Geography, Konkuk University, 05029, Republic of Korea

ABSTRACT

As global warming intensifies, the frequency and intensity of extremely high temperatures are expected to increase. This will impact the production of photovoltaics (PVs), which are increasingly adopted as an effective alternative to replace fossil fuel–based energy sources and reduce $CO_2$ emissions. Furthermore, extremely high temperature days account for a considerable proportion of days with high PV power potential (PVpot). Therefore, this study investigates changes in PVpot on future extremely high temperature days over East Asia, a region with high greenhouse gas emissions and vulnerability to extreme climatic events. The East Asia–averaged PVpot for extremely high temperature days was estimated to decrease across all scenarios and future periods. The East Asia–averaged PVpot for extremely high temperature days was predicted to decrease more substantially toward the late 21st century, with a larger magnitude of decrease expected under the high–carbon emissions scenario compared to the low–carbon emissions scenario. By the mid–and late 21st century, PVpot for extremely high temperature days was projected to decrease in PV hotspot areas, particularly in the regions of northern China and southern Mongolia, by up to −7.2 %. The signs of PVpot projections vary across sub–regions under summer mean conditions, while on extremely high temperature days, PVpot is consistently expected to decrease in all regions. This suggests that extremely high temperatures further intensify the decrease in PVpot. Moreover, under extremely high-temperature conditions, near-surface air temperature has been identified as the primary driver of projected decreases in PVpot among the climate variables considered; its influence is expected to intensify over time, thereby accelerating PVpot decreases under the high–carbon emissions scenario. Based on the findings, this study is expected to provide new insights for the development of renewable energy policies in a future where extremely high temperatures are projected to increase.

## 1. Introduction

Global carbon dioxide emissions have been increasing annually. A country-level analysis reveals that, in 2021, three major East Asian countries–China, Japan, and South Korea–ranked among the top 10 emitters. This is attributed to their large populations, urbanization, and high density of industrial zones. While carbon dioxide emissions in traditional industrial regions such as North America and Western Europe have been declining since the mid–2000s, emissions in East Asia have been increasing rapidly since the early 2000s. Notably, East Asia has the highest concentration of megacities, which serve as global carbon cycle hotspots due to substantial carbon dioxide emissions from electricity consumption, transportation, and residential and commercial buildings, all primarily driven by fossil fuel use (IGES, 2004; Nangini et al., 2017; Global Carbon Project, 2025).

Renewable energy offers an effective alternative to replace conventional fossil fuel–based energy sources and reduce carbon dioxide emissions. Furthermore, according to the United Nations Sustainable

* Corresponding author.
** Corresponding author.
*E-mail addresses:* dhcha@unist.ac.kr (D.-H. Cha), skmin@postech.ac.kr (S.-K. Min).

Development Goals (UNSDGs), the expansion of renewable energy adoption is essential for achieving SDG 7, which aims to ensure universal access to affordable, reliable, sustainable, and modern energy services for all (UN, 2015). Intergovernmental Panel on Climate Change (IPCC) stated that solar power generation is the sector with the highest potential contribution to net emissions reductions and is also the largest contribution from the low–cost options of the category of less than USD 20 per $tCO_2$-eq (IPCC et al., 2022b). Moreover, since 2010, the levelized cost of energy (LCOE) for the photovoltaic (PV) technology has dropped considerably, making PV not only economically competitive with fossil fuels but also, in many regions, a more cost–effective option than electricity generated from fossil fuel–based sources (IPCC et al., 2022b). Accordingly, at the United Nations Framework Convention on Climate Change (UNFCCC) Conference of the Parties 28 (COP28), held in the United Arab Emirates (UAE) in 2023, over 130 countries pledged to triple their renewable energy generation capacity by 2030 compared to current levels (UNFCCC, 2023).

Continued global warming is projected to increase both the frequency and intensity of heat waves (Easterling et al., 2000; Meehl and Tebaldi, 2004; Schär et al., 2004). Statistically, an increase in mean temperature shifts the probability distribution, remarkably raising the occurrence of extremely high temperatures (hereafter EHTs), such as heatwaves, which correspond to the right tail of the distribution (IPCC et al., 2014). In general, PV power potential (PVpot) is high on heat wave days because solar radiation is less affected by cloud cover, and precipitation events are rare. However, the increasing frequency and intensity of EHTs due to global warming and the decrease in panel efficiency attributed to these high temperatures could pose substantial obstacles to the future expansion of renewable energy use and the development of related policies (Park et al., 2022). Fig. 1 illustrates the annual summer peak electricity demand (MWh) for Jeju Special Self–Governing Province, the largest island in South Korea, located in its southernmost region with relatively high temperatures. The data is based on hourly electricity demand during the summer months (June, July, and August) from 2007 to 2021, provided by the Korea Power Exchange. In 2007, the annual peak electricity demand was 551 MWh; by 2021, it had nearly doubled to 1012 MWh. This increase in peak electricity is regarded as being primarily attributed to rising electricity consumption for cooling due to the increased frequency and intensity of EHTs during summer.

As the demand for transitioning from fossil fuel sources to renewable energy grows, the scale and output of solar power generation are expected to increase. Studies using climate models commonly projected that both the intensity and frequency of EHTs over East Asia will increase in the future compared to the present (Im et al., 2019; Park and Min, 2019; Wang et al.; Su and Dong, 2019; Xie et al., 2021; Kim et al., 2023). Therefore, during summer, when electricity demand is high, the efficiency of solar panels may be reduced, potentially leading to a decrease in power generation per unit area. In summary, the increasing frequency and intensity of EHTs in the future could lead to increased uncertainty and complexity in policy development related to renewable energy. Therefore, it is essential to provide predictive information on PV power generation based on future changes in EHTs under both low–carbon and high–carbon emission scenarios for East Asia, a region characterized by the highest carbon dioxide emissions and the most active renewable energy development.

As previously mentioned, East Asia is densely populated and has numerous industrial zones with substantial electricity consumption. As a result, the production of renewable energy, particularly in China, is currently the most active in the world. East Asian countries located in the mid–latitudes are expected to be highly vulnerable to extreme climates intensified by global warming (IPCC, 2012, 2013). Therefore, it is crucial to identify recent changes in PVpot and project future changes under conditions of EHTs, as temperature considerably affects the efficiency of solar panels. This study, as a follow–up to Park et al. (2022), which identified global warming as a major contributor to the projected decrease in future PVpot over East Asia, analyzes the impact of the increased frequency and intensity of EHTs caused by global warming on current PVpot over East Asia. Moreover, it projects these changes using high–resolution regional climate models (RCMs) covering the East Asian domain.

To achieve this, we ensembled high-resolution RCMs produced by the Coordinated Regional Climate Downscaling Experiment (CORDEX) project in the East Asian domain for a detailed spatiotemporal investigation of future PVpot changes focused on EHTs and applied them to our study. In this study, we used the Shared Socioeconomic Pathway (SSP) scenarios created to respond to IPCC AR6 (Sixth Assessment Report) by considering not only radiative forcing changes but also socioeconomic changes. High–resolution RCMs produced through the CORDEX project, which have the advantage of reliably capturing regional scale climate characteristics and providing more credible projections of future climate changes compared to GCMs, provide valuable information on future climate change and are widely utilized in various fields for climate change impact assessments. Recently, CORDEX RCMs have been increasingly applied to future potential projection studies on PV power generation, which is directly influenced by weather and climate at the East Asian scale (Park et al., 2022; Zhang et al., 2022; Ha et al., 2023; Kim et al., 2025). However, these studies have primarily projected future PVpot based on average climate conditions. Despite the vulnerability of solar panel efficiency to high temperatures, no study has investigated the impact of the increased frequency and intensity of EHTs over East Asia due to global warming on future PVpot. Thus, this study is the first to quantitatively assess the impact of projected future changes in extreme climate on renewable energy production using high–resolution RCMs over East Asia, a region characterized by the highest carbon dioxide emissions globally due to its concentrated population and industrial activity, and a critical need for the expansion of renewable energy generation.

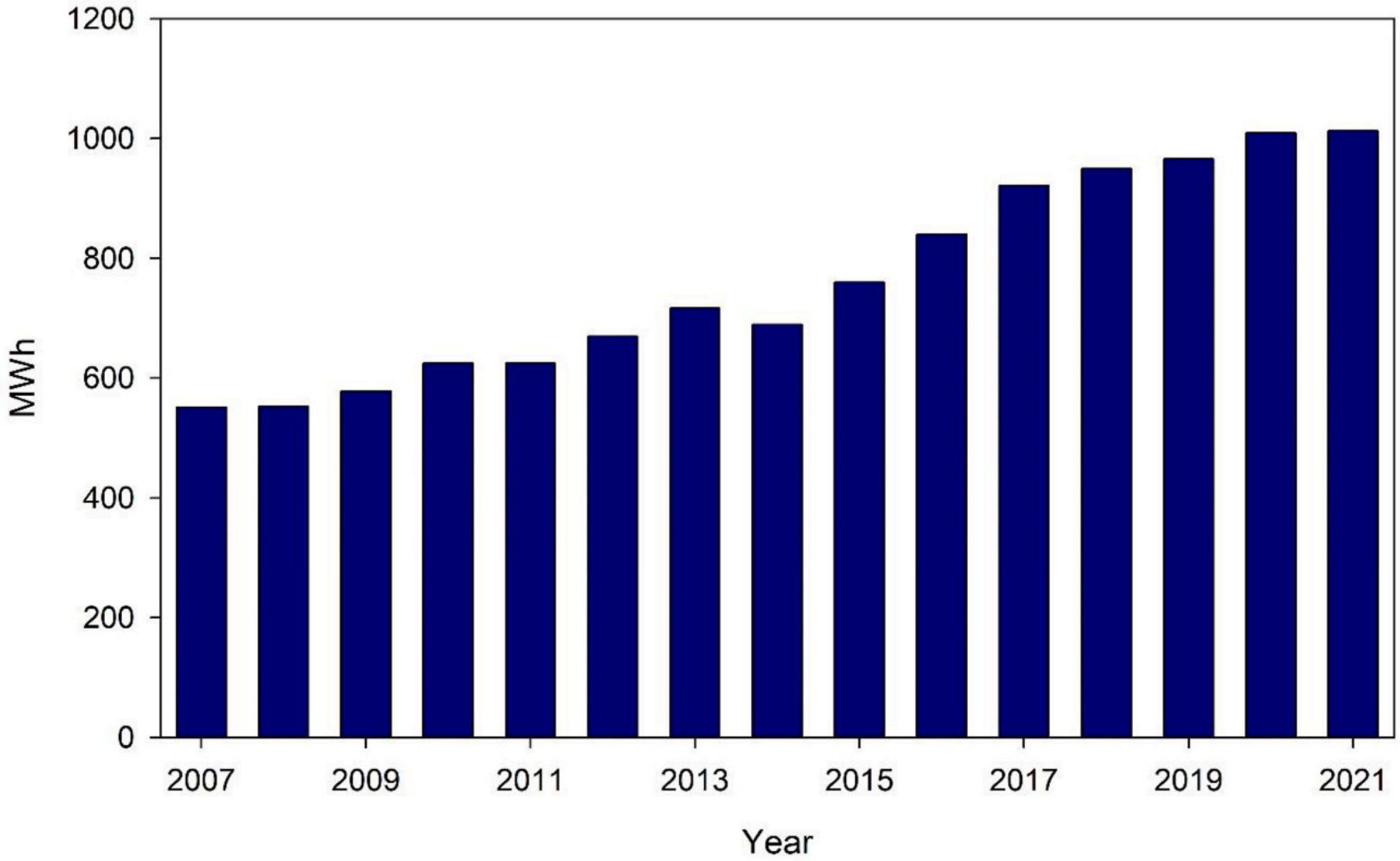

**Fig. 1.** The annual peak electricity demand (MWh) during the summer of 2007–2021 in Jeju Special Self–Governing Province of South Korea.

The remainder of this paper is organized as follows. Section 2 introduces the observational dataset, regional climate models, and analysis methods. Section 3 examines the observed PVpot for summer mean and EHT days over East Asia, and the projected future changes in PVpot of RCM MMEs for summer mean and EHT days over East Asia using RCMs produced through the CORDEX–East Asia Phase II project. Finally, Sections 4 and 5 present the summary and conclusion, and discussion, respectively.

## 2. Data and methods

### 2.1. Observational data and models

We selected the land regions of East Asia between 80 °E−150 °E and 20 °N–50 °N as the research domain, which includes China, the Korean peninsula, Japan, and parts of Mongolia and Russia (Fig. 2). The study area, located in the eastern part of the Eurasian continent, is almost identical to the East Asia domain, one of the six sub-regions of Asia as defined by the IPCC (2014; 2022a). This region is dominated by the East Asian Summer Monsoon (EASM) during the boreal summer, resulting in a rainy season (Chang, 2004). Rising temperatures have intensified the monsoon, thereby increasing the potential risk of flooding in this region. Moreover, the increasing frequency and intensity of EHTs and the urban heat island effect have amplified public health vulnerability in this densely populated region (IPCC et al., 2022a; World Bank, 2023). East Asia is one of the world's most dynamic regions, with a population density of approximately 140 people per square kilometer (World Bank, 2025), more than double the global average of 63 people per square kilometer (https://data.un.org/en/regions.html), and an economic growth rate of 5.0 % in 2024. Since a large proportion of the population resides in coastal megacities, East Asia is considered one of the regions most at risk of severe economic losses associated with sea–level rise (IPCC et al., 2022a; Hallegatte et al., 2013).

The major variables for the observational data and the RCMs used in this study include daily surface down–welling shortwave radiation (RSDS), daily near–surface air temperature (TAS), and daily near–surface wind speed (SFCWIND). RSDS directly influences the solar power generation, while TAS and SFCWIND are known to affect the efficiency of solar panels (Jerez et al., 2015; Bichet et al., 2019; Pérez et al., 2019). Daily maximum near–surface air temperature (TASMAX) has been used to indicate EHTs.

The European Centre for Medium-Range Weather Forecasts reanalysis 5 (ERA5) datasets with a horizontal resolution of 0.25° × 0.25° covering the period from 1979 to 2022 were utilized to evaluate the performance of the RCMs and investigate the recent changes in PVpot for the mean and extreme temperature conditions over East Asia. Additionally, future changes in East Asian PVpot for mean and extreme temperature conditions were estimated using five different RCMs with 25–km horizontal resolution forced by UKESM1 (UK's Earth System Model version 1) one of the CMIP6 (Coupled Model Intercomparison Project 6) GCMs (Global Climate Models), participating in the CORDEX–East Asia phase II project. The UKESM1 GCM was evaluated to have better simulation performance of major climate factors over East Asia than the CMIP5 GCMs (Shim et al., 2021; Sung et al.). Therefore, the UKESM1 GCM was applied as a suitable driving force for dynamical downscaling to produce high–resolution RCMs for East Asia under the SSP scenario through the CORDEX–East Asia phase II project. Table 1 outlines the configurations of five RCM simulations with different physical schemes applied in this study. The *Historical* experiment, spanning 36 years from 1979 to 2014, was used to evaluate the performance of the RCMs and applied as the reference dataset for future PVpot projections. Two emission scenarios, *SSP1–2.6* and *SSP5–8.5*, were used to estimate and compare future PVpot over East Asia under varying carbon emission levels for three future periods: near–term (2025–2049), mid–term (2050–2074), and long–term (2075–2099), each spanning 25 years.

The *SSP1–2.6* scenario represents a low–carbon emission pathway, emphasizing sustainability through the development of environmentally friendly technologies and the expansion of renewable energy use. Therefore, this brings about a future with low challenges in mitigating and adapting to climate change. In contrast, the *SSP5–8.5* scenario depicts a low challenge for climate change adaptation due to high levels of technological and economic growth. However, this scenario represents a high–carbon emission level because the energy system is still heavily dependent on fossil fuels. Therefore, climate change mitigation is expected to be a highly challenging path (O'Neill et al., 2014; Riahi et al., 2017). To reduce the uncertainties among RCMs caused by different model configurations, we constructed a multi–model ensemble (MME) by averaging the outputs of five RCMs with equal weighting. All five RCM outputs were post–processed using bilinear interpolation to match the 0.25° × 0.25° grid resolution of the ERA5 dataset.

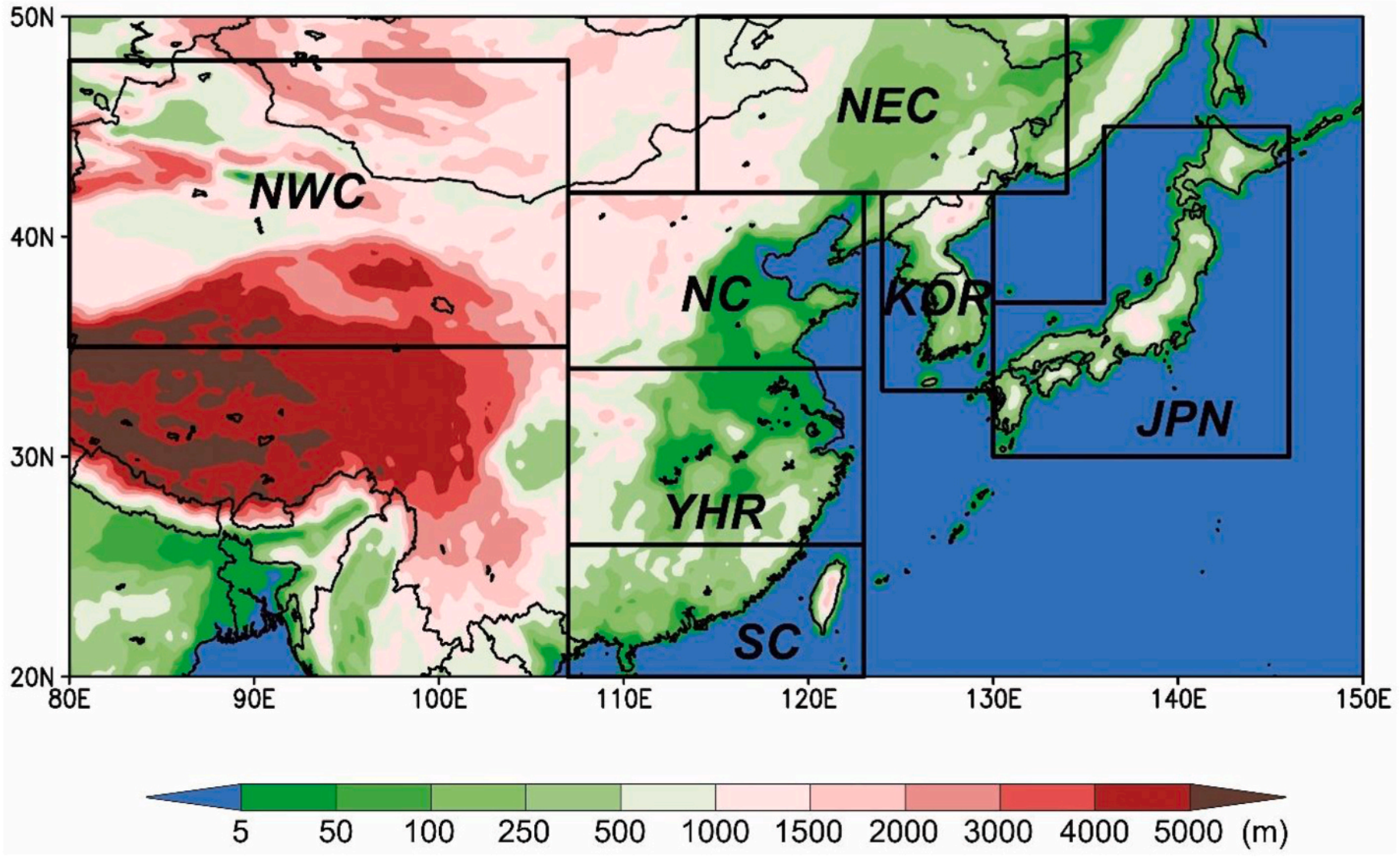

**Fig. 2.** Research domain and selected seven sub–regions over East Asia: Northwest China (*NWC*), Northeast China (*NEC*), North China (*NC*), Yangtze–Huaihe River Basin (*YHR*), South China (*SC*), Korean peninsula (*KOR*), and Japan (*JPN*). Shading indicates the altitude (Unit: Meter).

**Table 1**
Configurations of RCMs used in this study.

| Model | Resolution and number of grid points (lat. × lon.) | Vertical levels | Convection schemes | Microphysics | Radiation | Land surface model | References |
|---|---|---|---|---|---|---|---|
| HadGEM3-RA | 25 km, 251 × 396 | 63 eta | Revised mass flux | Single moment bulk | General 2–stream radiation | Joint UK Land Environment Simulator (JULES) | Davies et al. (2005) |
| WRF v.4 | 25 km, 250 × 395 | 30 eta | Betts–Miller–Janjic | WSM3 | Community Atmospheric Model radiation scheme (CAM) | NOAH | Gochis et al. (2017) |
| GRIMs | 25 km, 252 × 401 | 28 sigma | Simplified Arakawa–Schubert | WSM1 | LW: Chou, SW: Chou and Suarez | NOAH | Hong et al. (2013) |
| CCLM v.5 | 25 km, 231 × 376 | 40 hybrid | Extended DM | Extended DM | Ritter and Galeyn | TERRA ML | Doms and Baldauf (2013) |
| RegCM v.4 | 25 km, 249 × 394 | 23 sigma | MIT–Emanuel | SUBEX | NCAR CCM3 | NCAR CLM3.5 | Giorgi et al. (2012) |

### 2.2. Analysis methods

This study estimated PVpot over the East Asian region by applying the method suggested by Jerez et al. (2015), as in the previous study by Park et al. (2022). PVpot is an indicator of PV cell performance with a dimensionless magnitude. It is expressed as follows;

$$PV_{pot}(t) = P_R(t)\frac{RSDS(t)}{RSDS_{STC}} \quad (1)$$

where $RSDS_{STC}$ is defined as 1000 W m$^{-2}$. The performance ratio ($P_R$) is calculated as follows;

$$P_R(t) = 1 + \gamma[T_{cell}(t) - T_{STC}] \quad (2)$$

Here, $T_{STC}$ and $\gamma$ are set to 25 °C and −0.005 °C$^{-1}$, respectively. $T_{cell}$ represents the temperature of PV cell, which is estimated using three variables as follows;

$$T_{cell}(t) = 4.3\,^{\circ}C + (0.943 \cdot TAS(t)) + \left(0.028^{\circ}C\,m^2\,W^{-1} \cdot RSDS(t)\right) + \left(-1.528\,^{\circ}C\,s\,m^{-1} \cdot SFCWIND(t)\right) \quad (3)$$

TAS and SFCWIND–induced changes in PVpot change are estimated as follows;

$$\Delta TAS\ induced\ PV_{pot}(t) = \left(\frac{\alpha_1 RSDS(t) \cdot \Delta TAS}{PV_{pot_{Historical}}mean}\right) \cdot 100 \quad (4)$$

$$\Delta SFCWIND\ induced\ PV_{pot}(t) = \left(\frac{\alpha_2 RSDS(t) \cdot \Delta SFCWIND}{PV_{pot_{Historical}}mean}\right) \cdot 100 \quad (5)$$

where $\alpha_1 = -4.715 \times 10^{-6}$ and $\alpha_2 = 7.64 \times 10^{-6}$. The RSDS–induced changes in PVpot change were simply calculated by subtracting the $\Delta TAS\ induced\ PV_{pot}$ and $\Delta SFCWIND\ induced\ PV_{pot}$ from PVpot change. These values are represented in percentage form.

The Expert Team on Climate Change Detection and Indices (ETCCDI), organized by the World Meteorological Organization (WMO), developed 27 standardized core extreme indices for temperature and precipitation (Karl et al., 1999; Peterson et al., 2001; Frich et al., 2002; Peterson, 2005; Klein Tank et al., 2009; Sillmann et al., 2013). These indices are defined using either fixed thresholds or percentile–based thresholds. Notably, indices based on percentile thresholds offer the advantage of enabling objective and straightforward comparisons of extreme climate events between regions with different climate characteristics (Dunn and Morice, 2022). One such index, TX90P, is a percentile–based index defined as the percentage of days when the daily maximum temperature exceeds the 90th percentile of the daily values of the reference period (Zhang et al., 2005). In some studies, it has been applied in research by converting it to the number of days when the daily maximum temperature exceeds the 90th percentile (Dunn and Morice, 2022; Kim et al., 2022). The number of selected days each year may vary when applying the TX90P definition based on a base period to identify EHTs. Therefore, to obtain a consistent number of EHT days each year, the days with the daily maximum temperature exceeding the 90th percentile throughout the year were selected as the occurrence days of EHTs. The future projection of PVpot over East Asia under two carbon emission scenarios, *SSP1–2.6* and *SSP5–8.5*, was expressed as a percent (%) change relative to the climatology of the *Historical* experiment.

We selected seven sub–regions within the domain to analyze PVpot changes on future EHT days in detail for each region (Fig. 2). The locations and abbreviations of these sub–regions are provided in the figure's caption.

## 3. Results

### 3.1. Observed PVpot for summer mean and EHT days and RCMs' performances over East Asia

Fig. 3 illustrates the spatial distribution of the means by periods and the recent changes in PVpot estimated for summer mean and EHT days during 1979–2022 over East Asia, compared to the climatology. The summer climate over East Asia has undergone a notable shift since the late 1990s (Matsumura and Horinouchi, 2023; Lee et al., 2023). Accordingly, the study period was evenly divided into two halves based on the year 2000: the first (P1: 1979–2000) and the second half (P2: 2001–2022), each spanning 22 years. The difference between these two periods was then calculated to assess recent trends. Since most of EHT days over East Asia occur during summer, the climatological analysis for comparison was focused on the summer averages. Over the 44-year period, PVpot for EHT days observed over East Asia was higher than the summer mean across all regions and exhibited similar spatial distribution patterns. As presented in Section 1, high temperatures can reduce solar panel efficiency. However, on EHT days, PVpot tends to be higher because solar radiation, which directly influences PVpot, is less affected by cloud cover, and precipitation events are rare. The recent overall changes in summer mean temperature and the mean temperature of EHT days for East Asia were +0.75 and + 0.77 °C, respectively, indicating similar increases in both cases, but with little change in PVpot for either (Fig. S1 and Fig. 3). Although the final year of the analysis period in this study extends four years beyond that of the previous study (Park et al., 2022), the PVpot distribution shows no notable differences. With the exception of the western highlands of China, the northern China and southern Mongolia regions still exhibit high PVpot. These areas are hotspots for solar power generation, where large–scale solar power complexes have been constructed. PVpot for EHT days has recently increased in Korea, central China, South China, and Japan, while no notable change has been observed in PV hotspot areas.

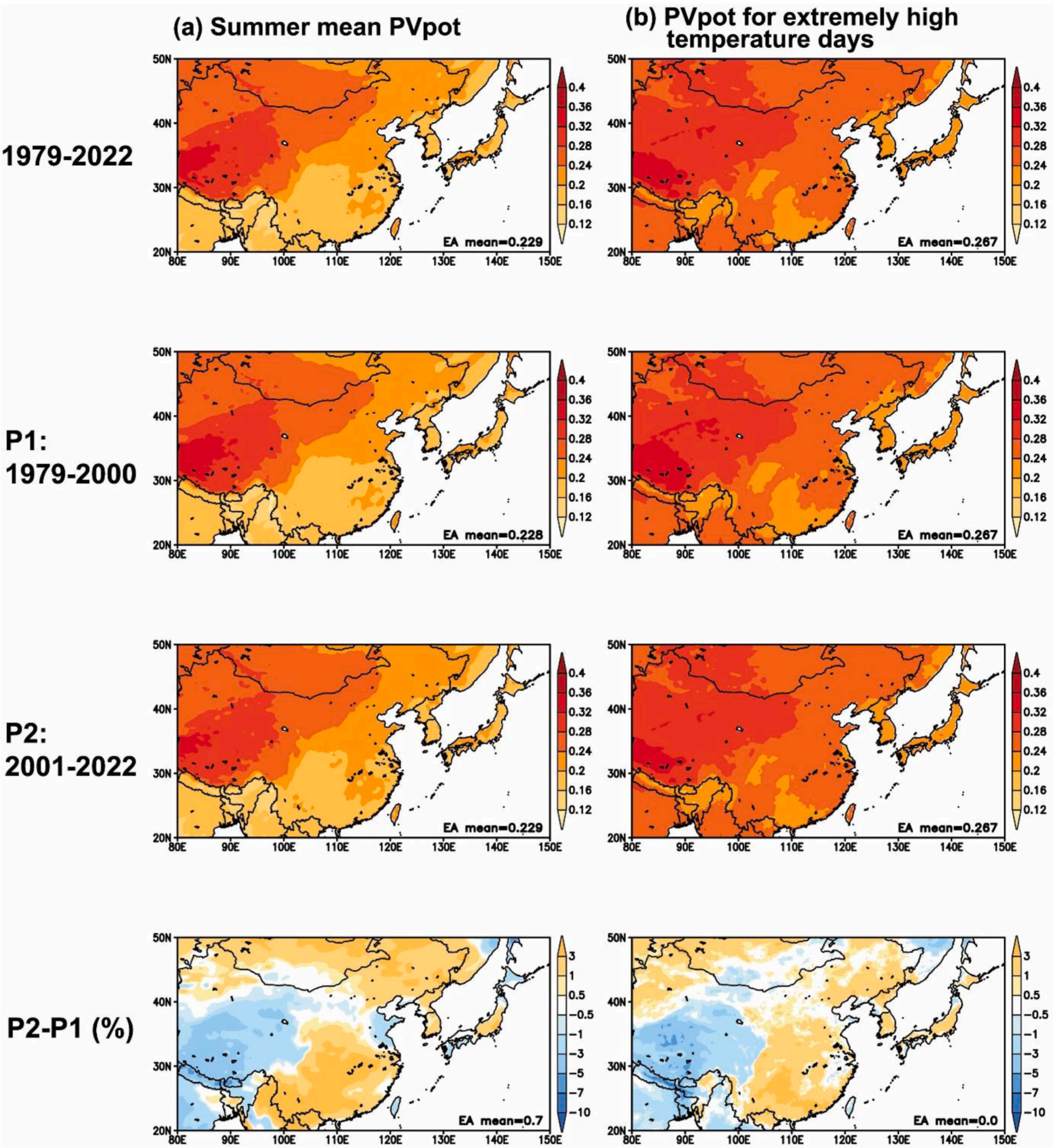

**Fig. 3.** Spatial distributions of the means by periods (the 1st to the 3rd columns) and the recent changes (the 4th column) in PVpot estimated for (a) summer mean and (b) EHT days over East Asia. The area–averaged values are provided in the bottom right corners.

We investigated the ratio of EHT days, defined as the days with the daily maximum temperature exceeding the 90th percentile of the year, among high PVpot days exceeding the 90th percentile of PVpot throughout the year (Fig. 4). The East Asia–averaged ratio of EHT days among high PVpot days during 1979–2022 was 30.6 % (Fig. 4(a)). This suggests that EHT days account for a substantial proportion of high PVpot days, and this result highlights the necessity of this study. In terms of overall spatial distribution, areas south of 30 °N latitude exhibited high ratios, while the Himalayan Plateau, the Korean Peninsula, and Japan showed relatively low ratios. The distribution of changes between the first and second halves of the analysis period revealed little regional variation, although a decrease of up to approximately 10 % was observed in the lower Yangtze River basin. The annual change in the East Asia–averaged ratio showed a statistically significant decreasing trend (Fig. 4(b)), suggesting that the recent increase in the mean temperature of EHT days has contributed to the decrease in the proportion of EHT days among high PVpot days.

Before projecting future changes in PVpot for summer mean and EHT days over East Asia using multi–RCM ensembles, we evaluated the performance of individual RCMs. First, PVpot biases between the RCM for each simulated *Historical* experiment and the ERA5 reanalysis data were examined (Fig. S2). All RCMs overestimated East Asia–averaged summer mean PVpot and PVpot for EHT days, but the magnitude of the overestimation was found to be quite small. HadGEM3–RA, CCLM, and RegCM4 showed low biases ranging from 0.01 to 0.03, with values similar to observations across most regions. GRIMs and WRF exhibited relatively higher biases, with notable overestimation in the western plateau of China. However, in the case of MME, the regional differences in PVpot overestimation were substantially reduced, and the bias was also smaller. The results of this study are subject to certain limitations, particularly uncertainties arising from model biases in the simulations and from inherent limitations in future projections. Although bias–correction techniques are sometimes used to mitigate such biases, they can also introduce additional uncertainties, requiring caution as they may lead to decisions that are not robust across plausible climate projections (Lafferty and Sriver, 2023); therefore, this study relies on the raw model outputs for its analysis.

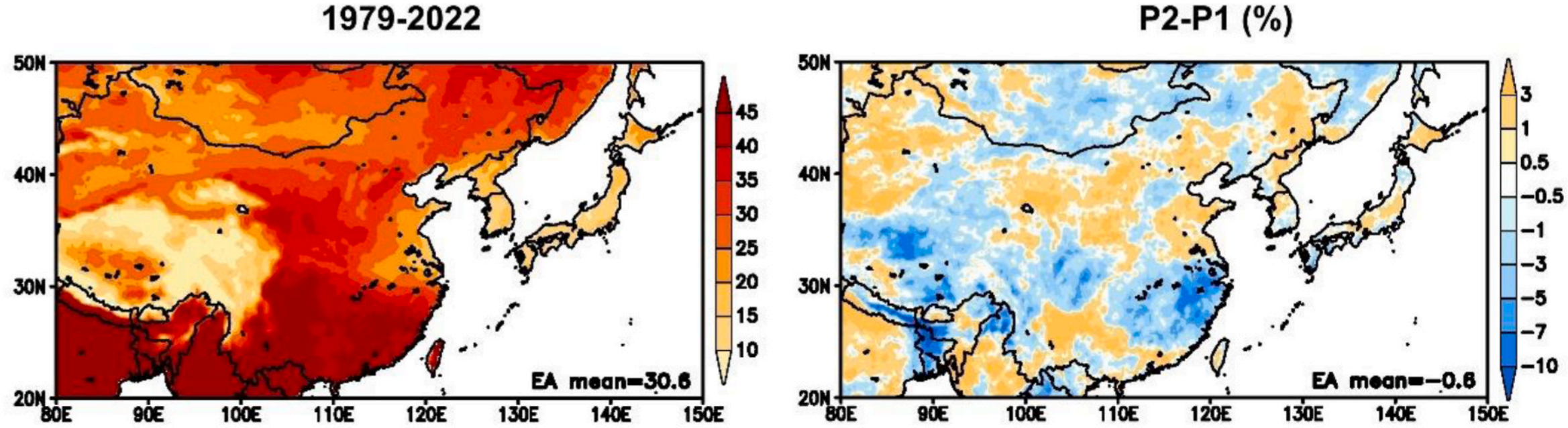

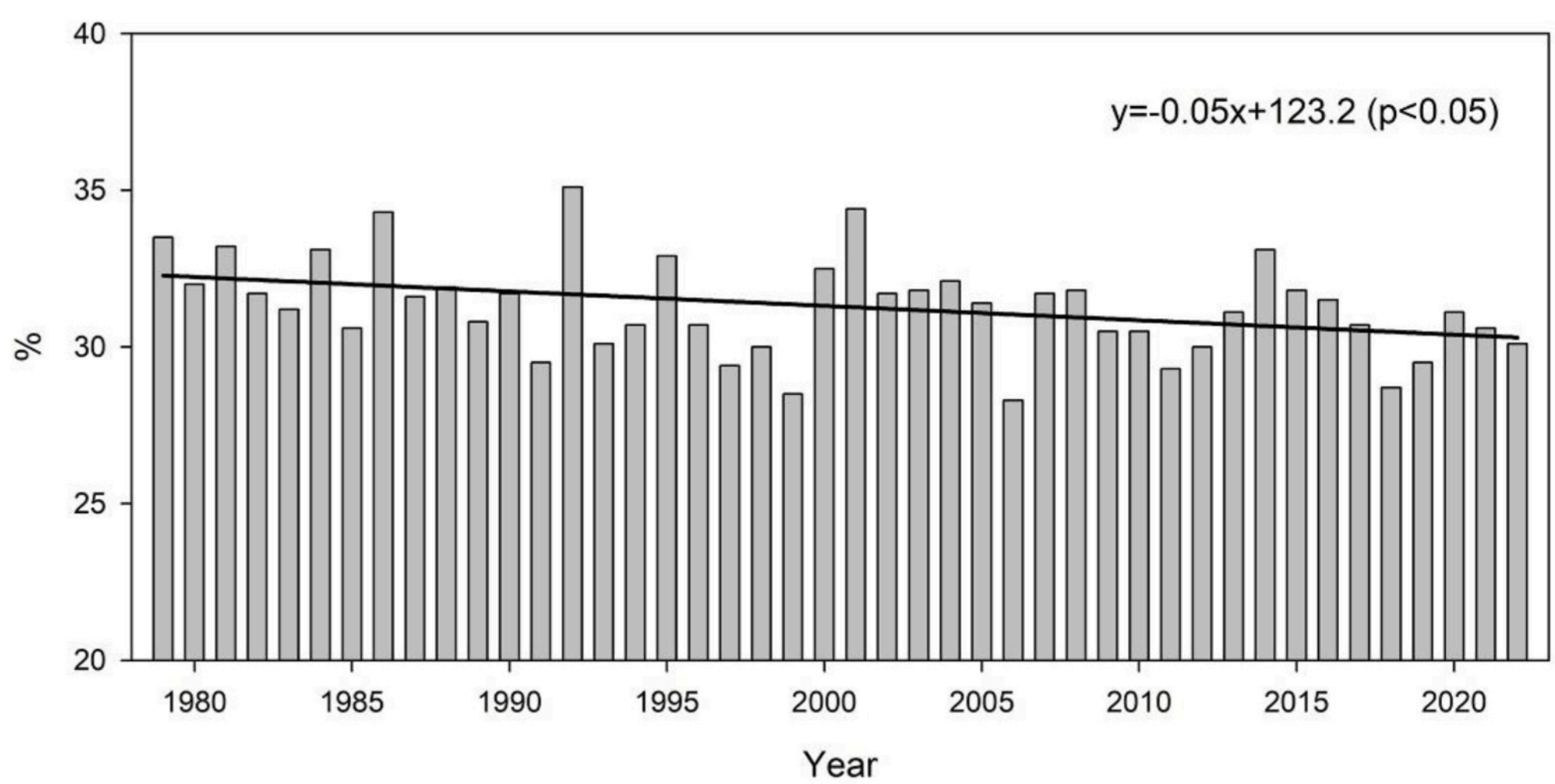

**Fig. 4.** (a) Spatial distributions for the ratio for EHT days among high PVpot days, and (b) annual time series of the East Asia–averaged ratio. In (a), The area–averaged values are provided in the bottom right corners.

Next, a Taylor diagram–widely used for evaluating climate model performance–was applied (Fig. 5). This method compares the standard deviation and correlation coefficient between the model and observation for spatial distribution (Taylor, 2001). A model located at the reference point in the diagram indicates perfect model performance, with the same spatial variability (standard deviation) as the observation and a spatial correlation coefficient of 1.0 with the observation. Thus, the distance from the RCM of each *Historical* experiment to the reference point reflects the model's error related to spatial distribution, and a shorter distance indicates a model with better performance. For summer mean PVpot, standard deviations ranged from 0.88 to 1.04 and correlation coefficients from 0.86 to 0.94. For PVpot for EHT days, standard deviations ranged from 1.02 to 1.41 and correlation coefficients from 0.72 to 0.92. The MME had standard deviations of 1.03 and 1.04 and correlation coefficients of 0.97 and 0.92 for summer mean PVpot and PVpot during EHT days, respectively. This indicates that the MME performed better than individual RCMs, with standard deviations closer to 1.0 and relatively higher spatial correlation coefficients, placing it nearer to the reference point. Based on these results, we will use the MME, which outperforms individual RCMs, to project future PVpot for EHT days over East Asia.

### *3.2. Future changes in PVpot of RCM MMEs for summer mean and EHT days over East Asia*

Using the *SSP1–2.6* low–carbon emission scenario and the *SSP5–8.5* high-carbon emission scenario, we projected future PVpot for EHT days over East Asia for the periods 2025–2049, 2050–2074, and 2075–2099. We also examined differences in PVpot for EHT days by emission scenario for each future period and compared summer mean PVpot with PVpot for EHT days. First, we examined the future changes in summer mean temperature and the mean temperature of EHT days over East Asia by future period and SSP scenarios (Fig. S3). All grids over East Asia will have positive changes in all future periods and under both scenarios, with larger values under high–carbon emission scenarios. These increases are projected to intensify toward the late 21st century, with the gap between scenarios widening over time. In both cases, the distributions by each future period and SSP scenario and East Asia–averaged future change values will be very similar to each other.

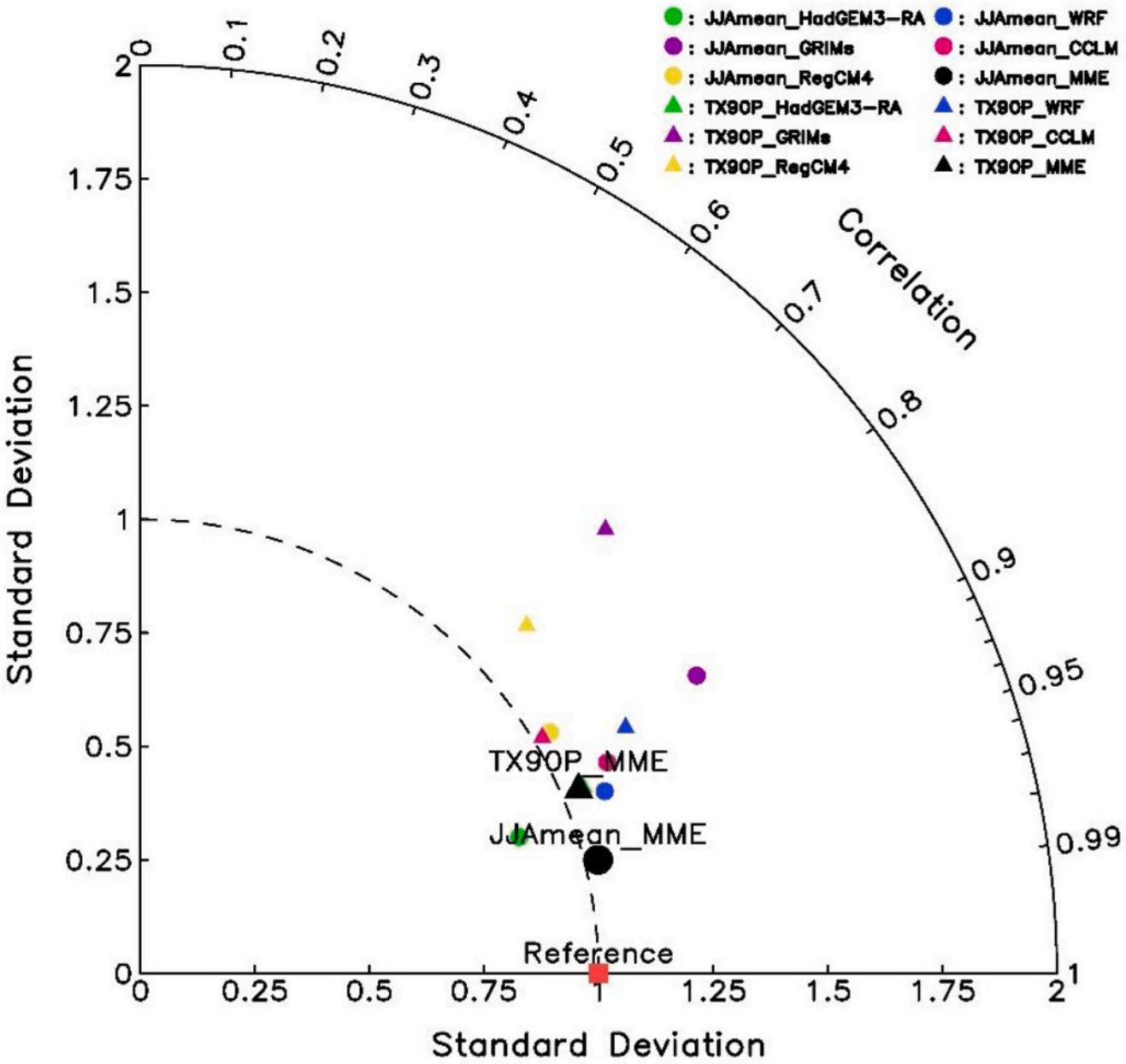

**Fig. 5.** Taylor diagram of PVpot for summer mean and EHT days of each RCM and MMEs from the *Historical* experiment over East Asia.

However, PVpot projections reveal notable differences. Fig. 6 depicts the spatial distributions of projected percentage changes in summer mean PVpot and PVpot for EHT days for each future period and scenario. Hatched areas indicate grids with strong inter–RCM agreement, defined as at least four of the five RCM simulations showing the same sign of change. East Asia–averaged summer mean PVpot and PVpot for EHT days, shown in the bottom–right corners of the figures, are expected to decrease under all scenarios and future periods. Notably, the spatial extent and magnitude of these decreases are expected to be larger for EHT days than for summer mean conditions. Moreover, PVpot is projected to decrease more substantially toward the late 21st century for both summer mean and EHT days, and the magnitude is projected to be larger under the high–carbon emissions scenario compared to the low-carbon emissions scenario. The decreases in PVpot for EHT days are especially noticeable in PV hotspot regions, like northern China and southern Mongolia, during the mid– and late 21st century. For southern China and the Yangtze River basin, summer mean PVpot is projected to increase overall, whereas PVpot for EHT days is expected to decrease. If these results relate to changes in RSDS, which directly affect solar power generation, then changes in RSDS will still have a major impact on the future summer average PVpot projections in southern China and the Yangtze River basin (Fig. S4). However, for EHT days, PVpot changes are predicted to remain negative despite positive RSDS projections. Therefore, the increasing frequency and intensity of future EHTs are expected to have a substantial impact on PVpot projection to the extent that they will exceed the influence of RSDS, a primary driver in PVpot projection, by drastically reducing solar panel efficiency.

We also analyzed future changes in the ratio of EHT days among high PVpot days–defined as those exceeding the 90th percentile of future PVpot over East Asia (Fig. 7). The East Asia–averaged ratio is projected to decrease toward the late 21st century, with larger decreases under the high–carbon emission scenario than the low–carbon emission scenario. Notably, a substantial decrease exceeding 10 % is expected in PV hotspot areas, particularly in northern China and southern Mongolia. Therefore, as shown in the results for the recent period in Fig. 4, the future increase in the mean temperature of EHT days is also expected to contribute to the decrease in the ratio of EHT days among high PVpot days. These findings, on EHT days that account for a substantial share of high PVpot days, provide additional evidence that the intensification of EHTs under global warming will play a dominant role in future decreases in PVpot.

We examined the absolute contributions of RSDS, TAS, and SFCWIND as climate factors influencing future PVpot changes on EHT days. Fig. 8 presents the absolute contributions of RSDS, TAS, and SFCWIND to the East Asia–averaged projection of future PVpot for three future periods and two SSP scenarios, calculated using Eqs. (4) and (5) in Section 2. The left and right panels correspond to the low–carbon and high–carbon emission scenarios, respectively. The projected decrease in PVpot (brown bar) for both summer mean and EHT days becomes larger toward the late 21st century, with a more substantial decrease under the high–carbon emission scenario. Across all scenarios and future periods, the absolute contributions of SFCWIND (green bar) to the East Asia–averaged projections of future PVpot are almost negligible. For both summer mean and EHT days, TAS contributions (red bar) will have the largest influence on the East Asia–averaged projections of future PVpot. As the negative contribution of TAS increases toward the late 21st century, the decrease in projected PVpot will also become larger. Likewise, since the negative contribution of TAS is larger under the high–carbon emission scenario compared to the low–carbon emission scenario, the projected magnitude of PVpot decrease is larger under the high–carbon emission scenario. The extent of the decrease in TAS contribution in both cases is predicted to remain similar across all periods and scenarios. For summer mean, RSDS contributions (cyan bar) lead to a positive change under the high–carbon emission scenario across all future periods, partially offsetting TAS–induced decreases in projected PVpot. However, the larger negative impact of TAS is expected to dominate, resulting in an overall decrease in PVpot. In summary, these results highlight that TAS plays the dominant role in contributing to future decreases in PVpot, even on EHT days, whereas the contributions of RSDS and SFCWIND are comparatively slight. This finding suggests that the efficiency loss associated with rising temperatures is the prevailing mechanism that controls PV performance.

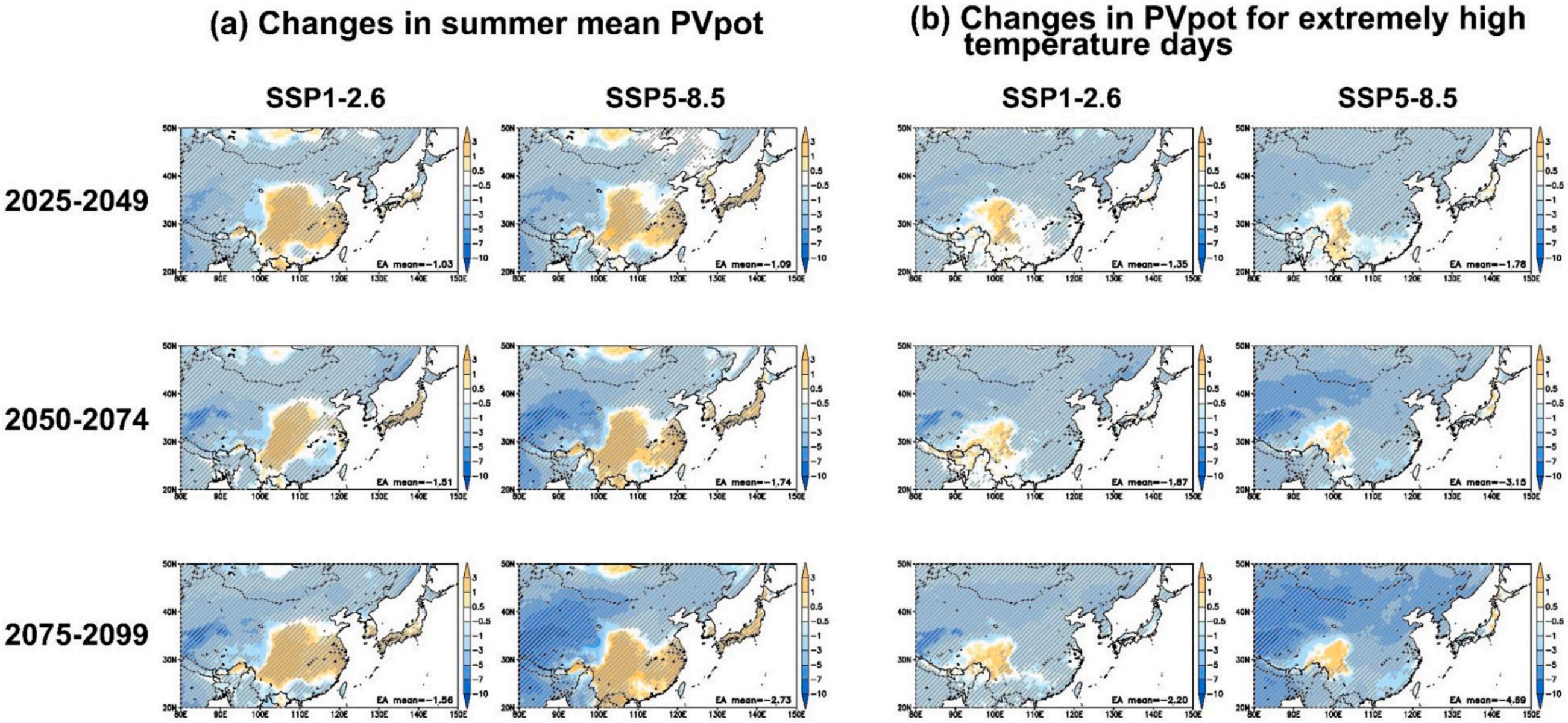

**Fig. 6.** Spatial distributions of future changes (%) in (a) summer mean PVpot and (b) PVpot for EHT days from three future periods and two SSP scenarios over East Asia. The area–averaged values are provided in the bottom right corners.

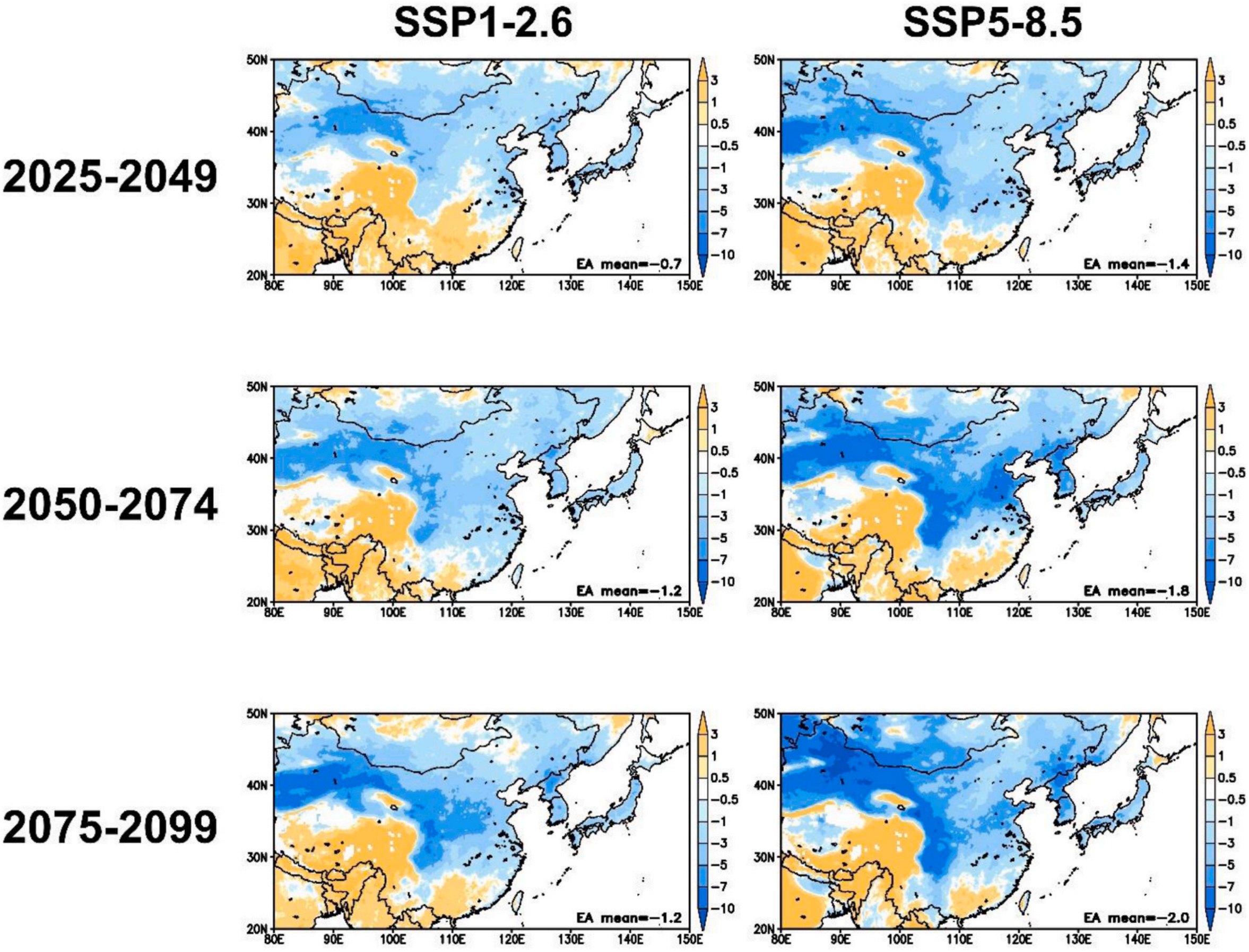

**Fig. 7.** Spatial distributions of the future change in the ratio of EHT days among high PVpot days for three future periods and two SSP scenarios over East Asia. The area–averaged values are provided in the bottom right corners.

Next, we examined the relative contribution (%) of each factor to the total contribution of all factors to East Asia–averaged PVpot changes across different future periods and SSP scenarios (Fig. 9). Plain bars indicate the summer mean, while the hatched bars correspond to EHT days. For summer mean, RSDS (cyan bar) has an opposing effect on PVpot decrease except for the 2050–2074 period under the low–carbon emission scenario. In contrast, TAS (red bar) is projected to be the largest contributor to PVpot decreases for both summer mean and EHT days across all future periods and SSP scenarios. Under the low-carbon emission scenario, the relative contribution of TAS to PVpot changes during EHT days is projected to decrease toward the late 21st century. This is because, as warming stabilizes after the mid–21st century, the contribution of RSDS to the decrease in PVpot on EHT days becomes relatively larger than that of TAS.

PVpot changes on future EHT days, classified into sub–regions of East Asia, are presented in Fig. 10. The left and right panels correspond to the low–carbon and high–carbon emission scenarios, respectively. Substantial regional differences are evident in the projections of summer mean PVpot over East Asia. The Yangtze River Basin region (*YHR*) is expected to experience an increase of PVpot across all future periods and scenarios, with a similar trend in the Japan region (*JPN*), except for the 2025–2049 period under the low–carbon emission scenario. Conversely, the *NWC* region, where the current PV hotspot for renewable energy production is located, is projected to experience the largest decrease among all sub–regions. This result is particularly concerning, as it suggests that the key area currently containing large–scale solar power complexes may face the highest future vulnerability, raising questions about the long–term sustainability of existing investment strategies for renewable energies. Thus, these findings highlight the need to consider such projections when establishing policies related to solar power generation in response to future warming in this region. For EHT days, PVpot is projected to decrease across all seven sub–regions under all future periods and scenarios. This decrease is expected to intensify toward the late 21st century, with a particularly pronounced impact under the high–carbon emission scenarios. Overall, the spatial pattern of decreases follows a "higher in the south, lower in the north" distribution, with the largest decrease occurring in the *NWC* region, as in the case of summer mean. Notably, under the high–carbon emission scenario during the 2075–2099 period, the decrease in the *NWC* region is projected to reach −7.2 %. The *YHR* region, where PVpot was expected to increase under summer mean conditions, will experience a decrease on EHT days. In summary, as previously shown, the future projections of summer mean temperature and the mean temperature of EHT days exhibit a similar increasing pattern (Fig. S3). However, the signs of PVpot projections vary across sub–regions under the summer mean condition, whereas for EHT days, PVpot is consistently expected to decrease in all regions. This apparently indicates that EHTs further amplify the decrease in PVpot, and emphasizes once again the necessity of this study in relation to the response to EHT in future PV policies.

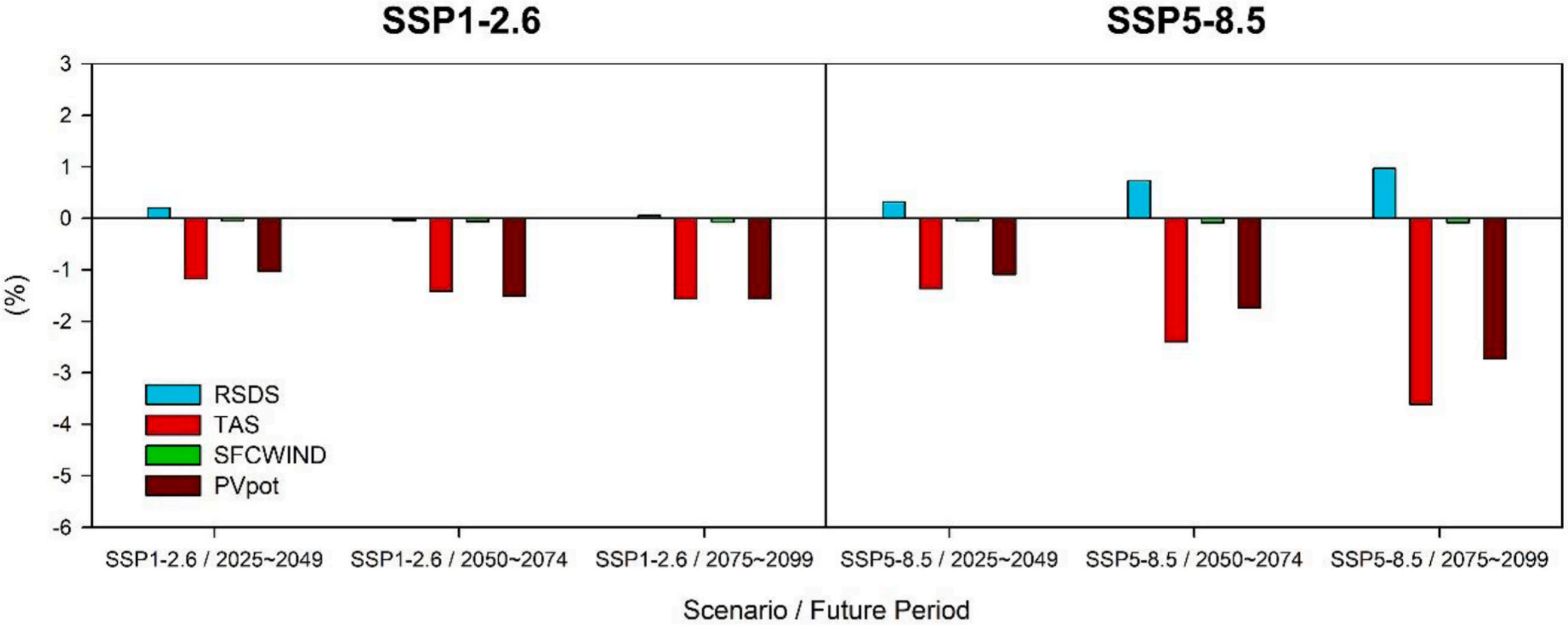

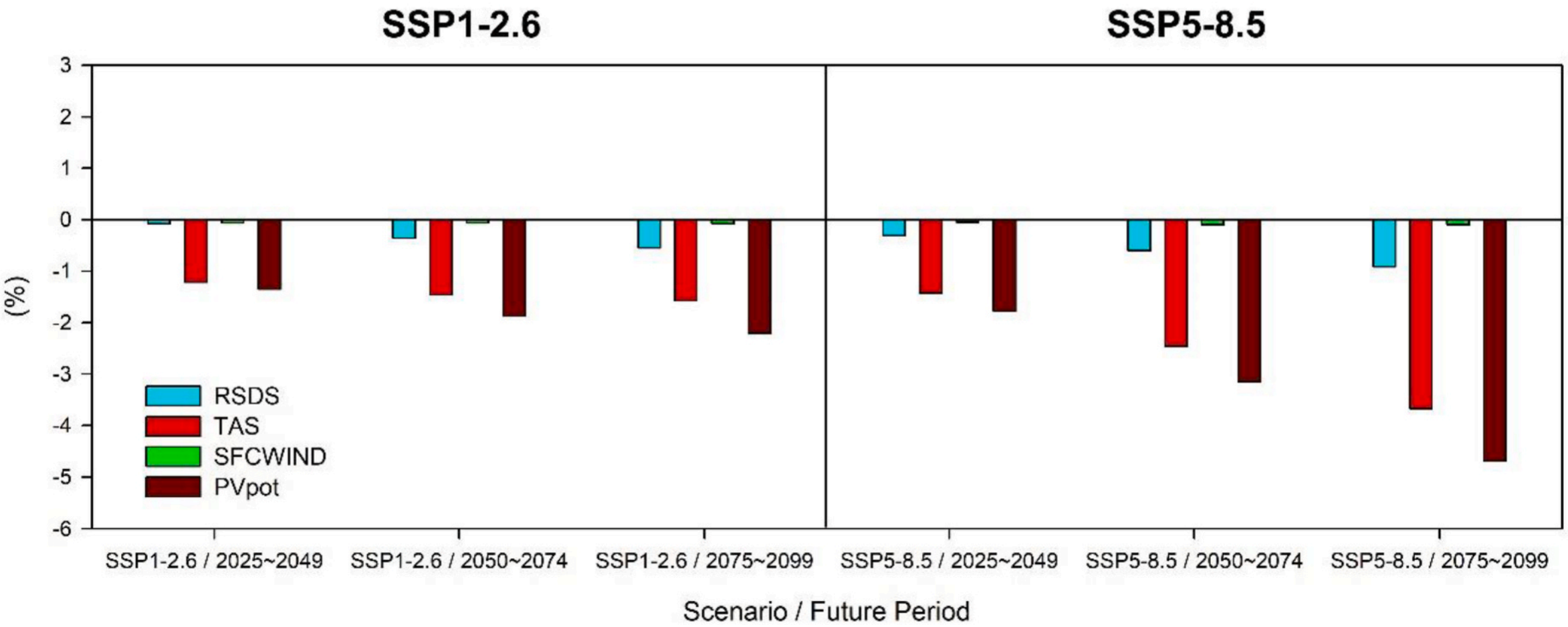

**Fig. 8.** Contributions of RSDS, TAS, and SFCWIND to the East Asia–averaged projection of future PVpot across three future periods and two SSP scenarios for (a) summer mean and (b) EHT days.

## 4. Discussion

As a follow–up to our previous study (Park et al., 2022), this research, applying new scenarios, further strengthens the conclusion that global warming could reduce future PVpot over East Asia. The expansion of PV power generation is essential for replacing fossil fuel–based energy sources and reducing carbon dioxide emissions. However, this study revealed that EHT days, which account for a substantial proportion of high PVpot days, are projected to increase in frequency and intensity due to global warming, leading to a reduction in PVpot during these days. In particular, the increase in frequency and intensity of EHT days will substantially increase electricity consumption for cooling, but PV power generation per unit area will decrease. Thus, despite the ongoing expansion of PV power generation, this paradoxical outcome could introduce larger uncertainty and complexity in renewable energy policies.

This study is the first to project the impact of future changes in East Asian EHT on PVpot using high-resolution RCMs based on SSP scenarios that consider not only scientific mechanisms but also future socioeconomic changes. Accordingly, this study can play a crucial role in reducing uncertainty or complexity in renewable energy policy and providing new insights. South Korea and Japan generate a lower percentage of electricity from renewable sources compared to the Organization for Economic Co–operation and Development (OECD) average. South Korea, in particular, recorded the lowest ratio, at approximately 6–7 % (IEA, 2021; The Korea Times, 2023; RE100, 2024). Accordingly,

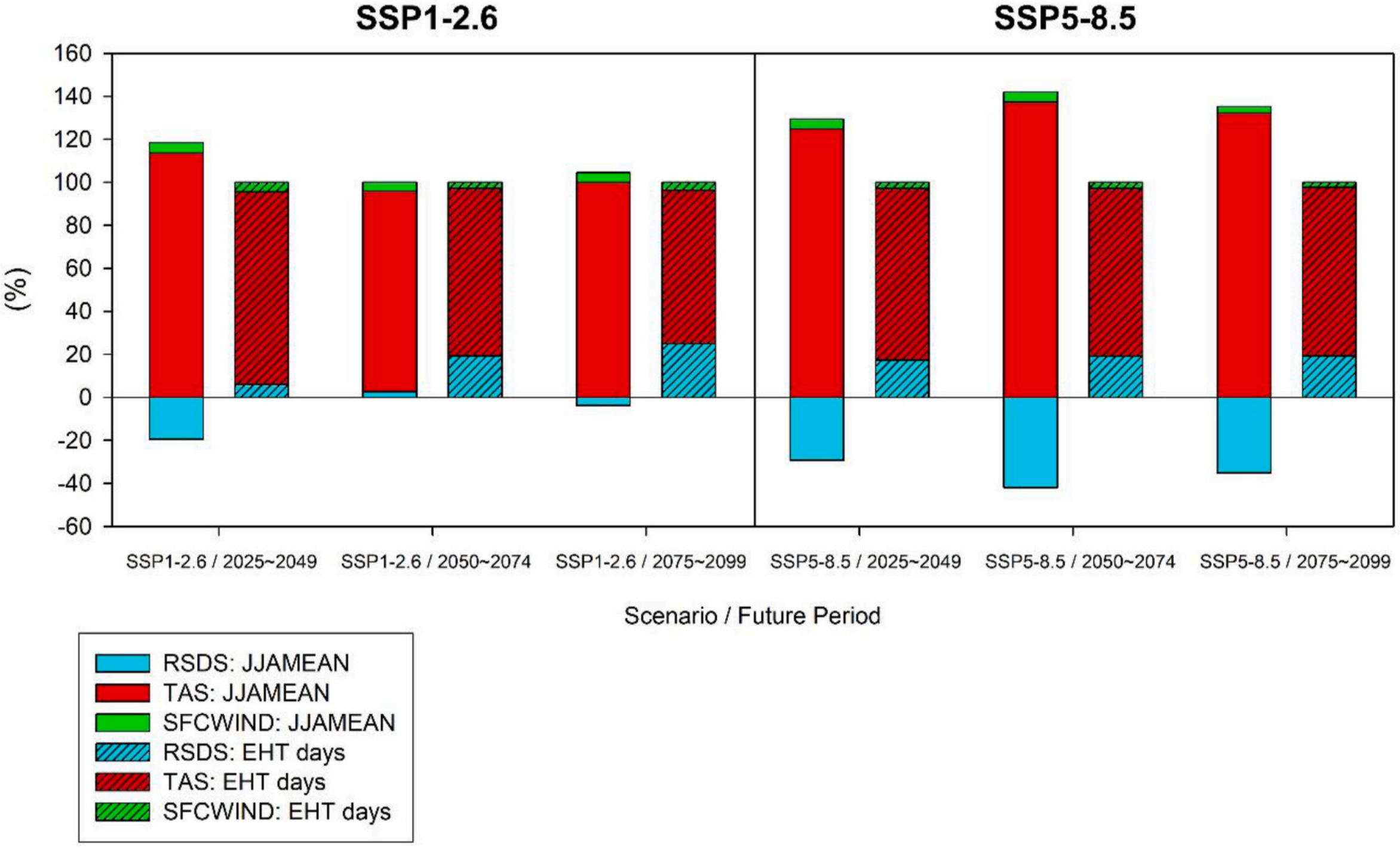

**Fig. 9.** Contributions of each variable to the East Asia–averaged PVpot change in percentage across three future periods and two SSP scenarios. Plain bars represent the summer mean, while the hatched bars correspond to EHT days.

Renewable Electricity 100 (RE100) has pointed out that South Korea's failure to take urgent and decisive measures to expand renewable energy will hinder South Korea's economic potential (RE100, 2022). Because East Asia accounts for the largest share of global $CO_2$ emissions, decisive policy action to accelerate the transition to renewable energy is essential. Moreover, as the frequency and intensity of EHT, which strongly influence PV efficiency, continue to rise across the region, the PV–EHT relationship established in this study can inform the development of electricity generation and supply management policies that balance conventional and renewable energy sources, particularly to prepare for and mitigate blackout risks associated with abrupt peaks in summer electricity demand in East Asia.

## 5. Summary and conclusion

This study examined the impact of the increased frequency and intensity of EHTs induced by global warming on the current PVpot and projected these changes using high–resolution RCMs under the multi–SSP scenarios across East Asia. Over the 44 years, PVpot associated with EHT days across East Asia was higher than the summer mean in all regions. In recent years, PVpot for EHT days has increased in Korea, central China, South China, and Japan, while the PV hotspot areas exhibited no notable changes. The recent increases in the mean temperature of EHT days across East Asia have contributed to a decrease in the proportion of EHT days among high PVpot days.

For future projections, the East Asia–averaged summer mean PVpot and PVpot for EHT days are expected to decrease across all scenarios and future periods. Furthermore, PVpot is projected to decrease more substantially toward the late 21st century for both summer mean and EHT days, with a larger magnitude of decrease expected under the high–carbon emissions scenario compared to the low–carbon emissions scenario. Overall, the decreased distribution is predicted to follow a "higher in the south, lower in the north" pattern, and in particular, by the mid–and late 21st century, PVpot for EHT days is projected to decrease substantially in PV hotspot areas in the regions of northern China and southern Mongolia, by up to −7.2 %. This indicates that the region is expected to face the greatest vulnerability in terms of future solar power generation, emphasizing the need for careful consideration of the long–term sustainability of renewable energy investment strategies in current PV hotspot region. Among the climate variables considered, TAS is identified as the primary driver of the projected decrease in PVpot during EHT days over East Asia. As the negative contribution of TAS is expected to intensify toward the late 21st century, especially under the high–carbon emission scenario, future PVpot projections were predicted to decrease over East Asia. These results suggest that the efficiency loss caused by rising temperatures represents the primary mechanism controlling PV performance, and climate warming is increasingly affecting solar energy production.

This study still has limitations, such as the insufficient spatial resolution of the applied RCMs and the constraint that only a single GCM was used as the forcing data for multiple RCMs. Nevertheless, the CORDEX-East Asia phase II project has taken the lead in producing RCMs based on SSP scenarios, and this study, which maximally employs the available RCMs, offers advanced findings for the development of renewable energy policies over East Asia, a region where renewable energy utilization is highly needed. In upcoming research, we plan to conduct a quantitative assessment of uncertainties in PVpot projections based on multiple scenarios and climate models. Additionally, we intend to examine the influence of natural climate variability on PVpot through atmospheric science-based analysis.

## CRediT authorship contribution statement

**Changyong Park:** Writing – review & editing, Writing – original draft, Visualization, Validation, Software, Resources, Methodology, Investigation, Funding acquisition, Formal analysis, Data curation, Conceptualization. **Ana Juzbašić:** Writing – review & editing, Software, Investigation, Data curation. **Dong-Hyun Cha:** Writing – review & editing, Writing – original draft, Supervision, Funding acquisition, Conceptualization. **Seung-Ki Min:** Writing – review & editing, Writing – original draft, Supervision, Conceptualization. **Joong-Bae Ahn:** Writing – review & editing, Data curation. **Eun-Chul Chang:** Writing – review &

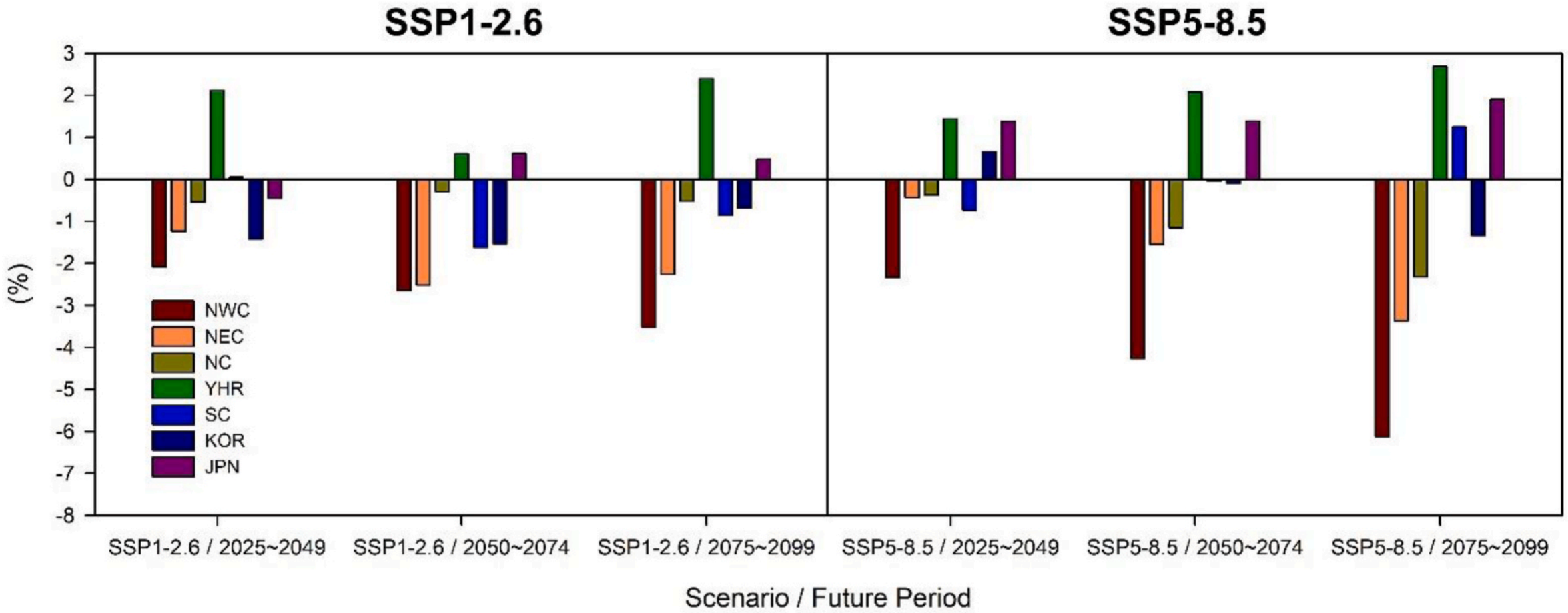

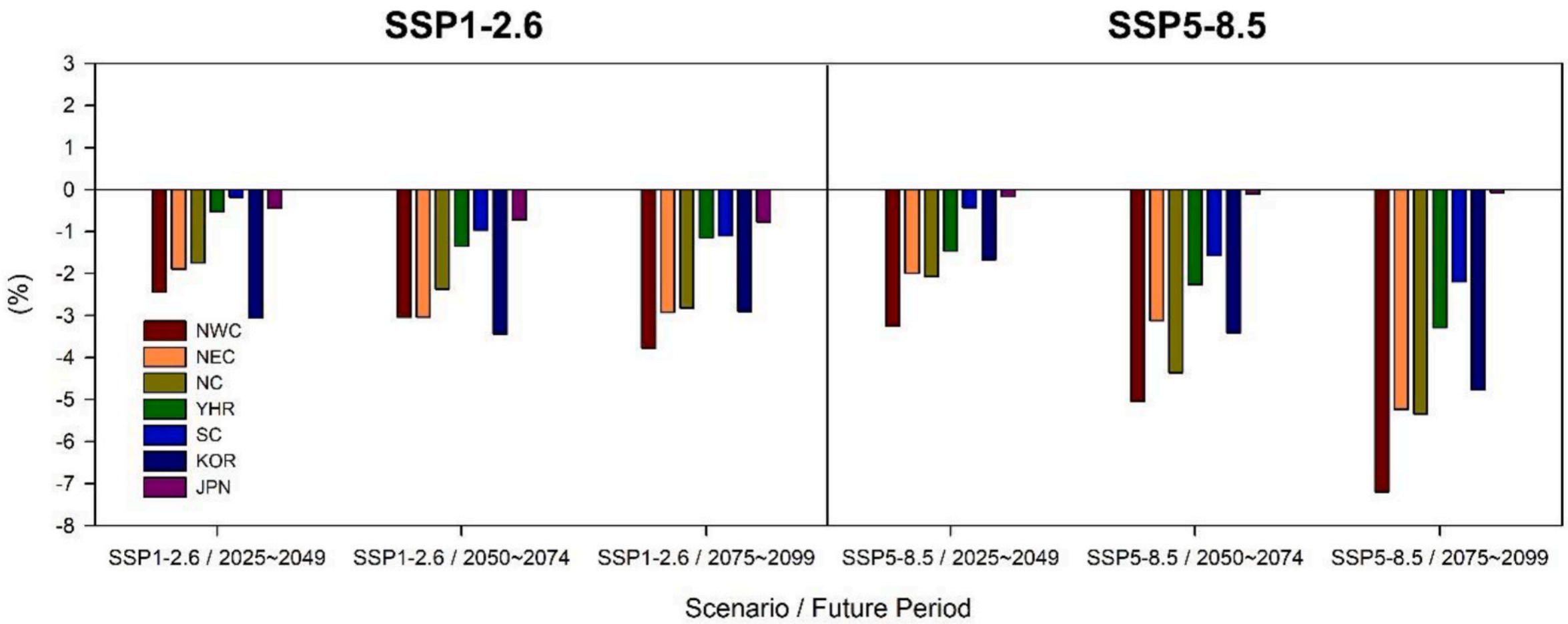

**Fig. 10.** Future changes in (a) summer mean PVpot and (b) PVpot for EHT days in detail by classifying East Asia into sub–regions across three future periods and two SSP scenarios.

editing, Data curation. **Young-Hwa Byun:** Writing – review & editing, Data curation. **Youngeun Choi:** Writing – review & editing, Data curation.

## Declaration of competing interest

The authors declare that they have no known competing financial interests or personal relationships that could have appeared to influence the work reported in this paper.

## Acknowledgments

This work was supported by the Korea Meteorological Administration Research and Development Program under Grant RS-2024-00403386 and in part funded by the National Research Foundation of Korea(NRF) under grant the Korea government(MSIT) (RS-2024-00352097).

## Appendix A. Supplementary data

Supplementary data to this article can be found online at https://doi.org/10.1016/j.wace.2026.100859.

## Data availability

Data will be made available on request.

## Supplementary material

# Impact of extremely high temperature on future photovoltaic power potential over East Asia

**Changyong Park[a], Ana Juzbašić[a], Dong–Hyun Cha[a,*], Seung–Ki Min[b,*], Joong-Bae Ahn[c], Eun-Chul Chang[d], Young–Hwa Byun[e], Youngeun Choi[f]**

[a] Department of Civil, Urban, Earth, and Environmental Engineering, Ulsan National Institute of Science and Technology, 44919, Republic of Korea

[b]Division of Environmental Science and Engineering, Pohang University of Science and Technology, 37673, Republic of Korea

[c]Department of Atmospheric Sciences, Pusan National University, 46241, Republic of Korea

[d]Department of Atmospheric Science, Kongju National University, 32588, Republic of Korea

[e]Department of Atmospheric Sciences, Yonsei University, 03722, Republic of Korea

[f]Department of Geography, Konkuk University, 05029, Republic of Korea

April 2025

*Corresponding authors

E–mail addresses: dhcha@unist.ac.kr (Dong–Hyun Cha) and skmin@postech.ac.kr (Seung–Ki Min)

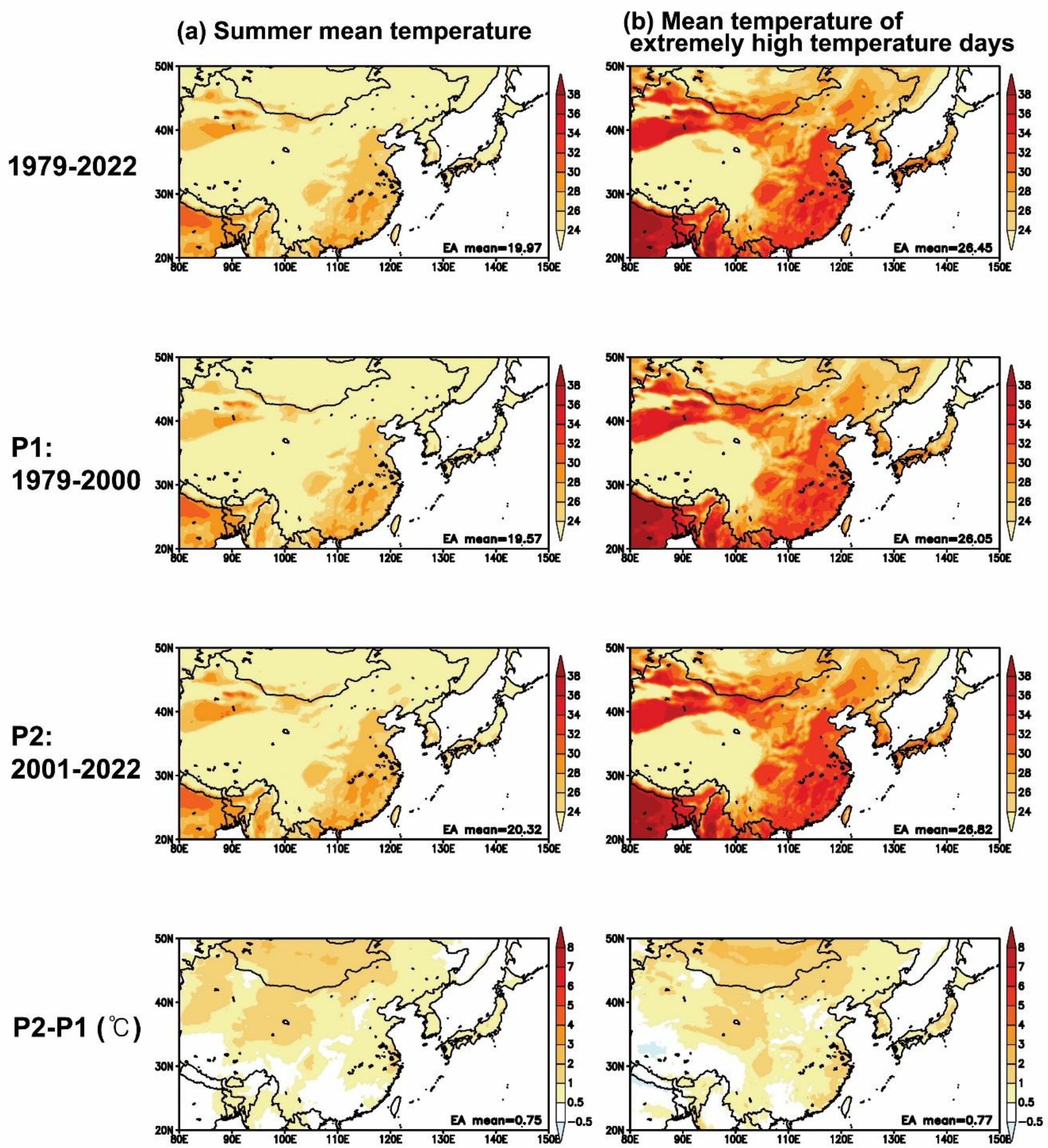

**Fig. S1.** Spatial distributions of the means by periods (the 1st to the 3rd columns) and the recent changes (the 4th column) in (a) summer mean temperature and (b) the mean temperature of EHT days over East Asia. The area–averaged values are provided in the bottom right corners.

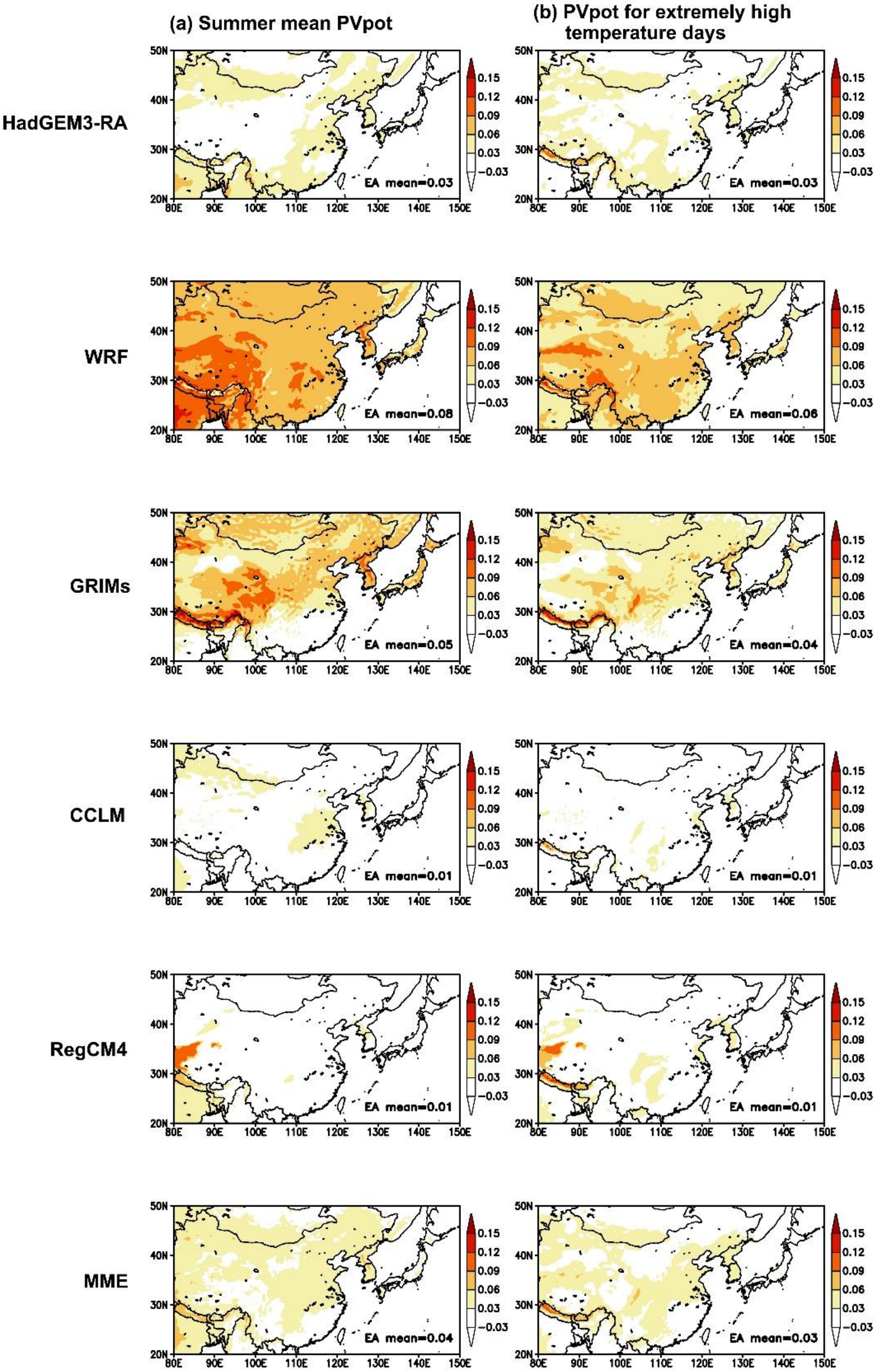

**Fig. S2.** Spatial distributions of PVpot bias between the RCM for each simulated *Historical* experiment and the ERA5 reanalysis data for (a) summer mean PVpot and (b) PVpot for EHT days. The area–averaged values are provided in the bottom right corners.

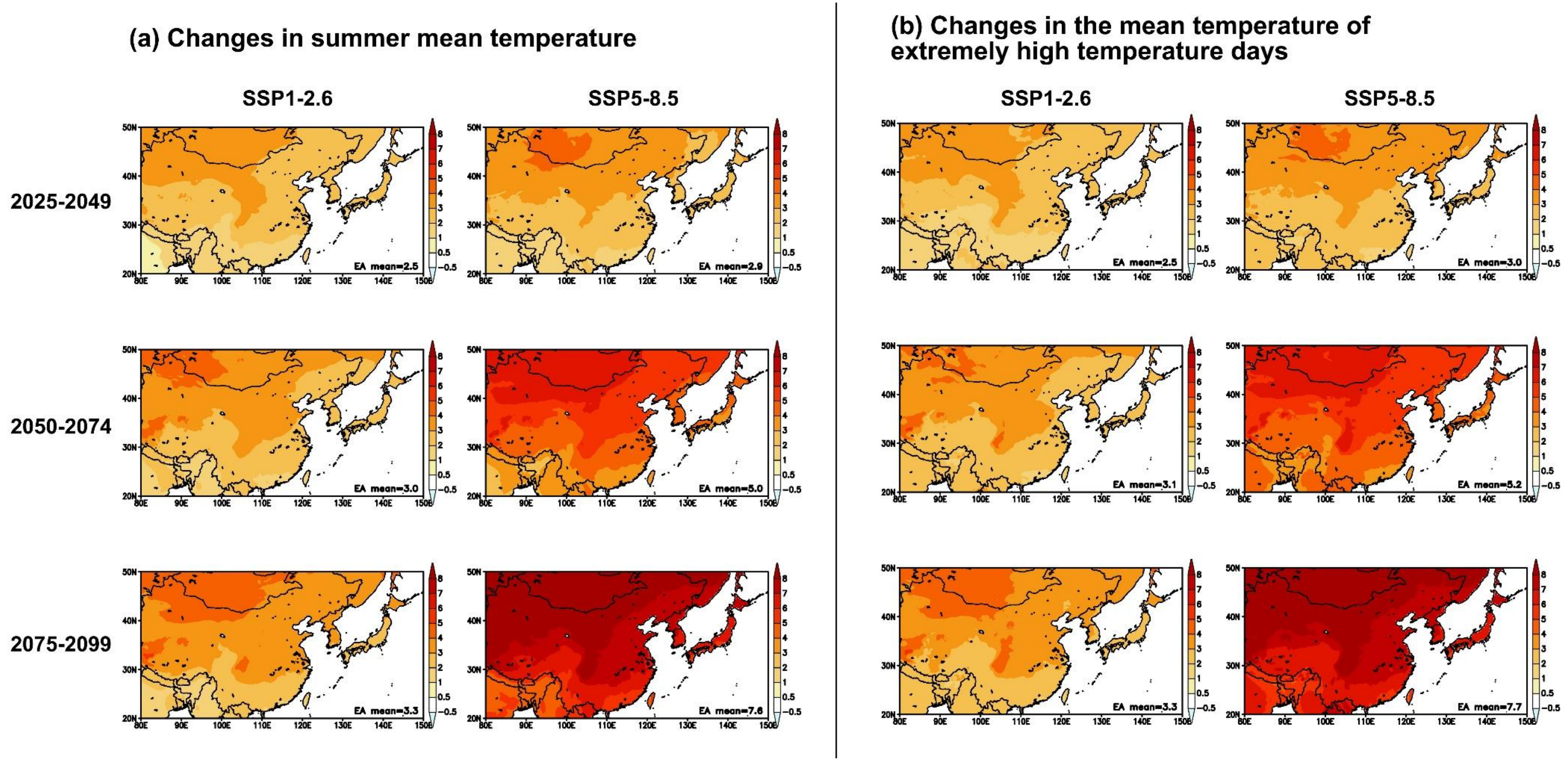

**Fig. S3.** Spatial distributions of the future changes (%) in (a) summer mean temperature and (b) the mean temperature of EHT days over East Asia by future period and SSP scenarios. The area–averaged values are provided in the bottom right corners.

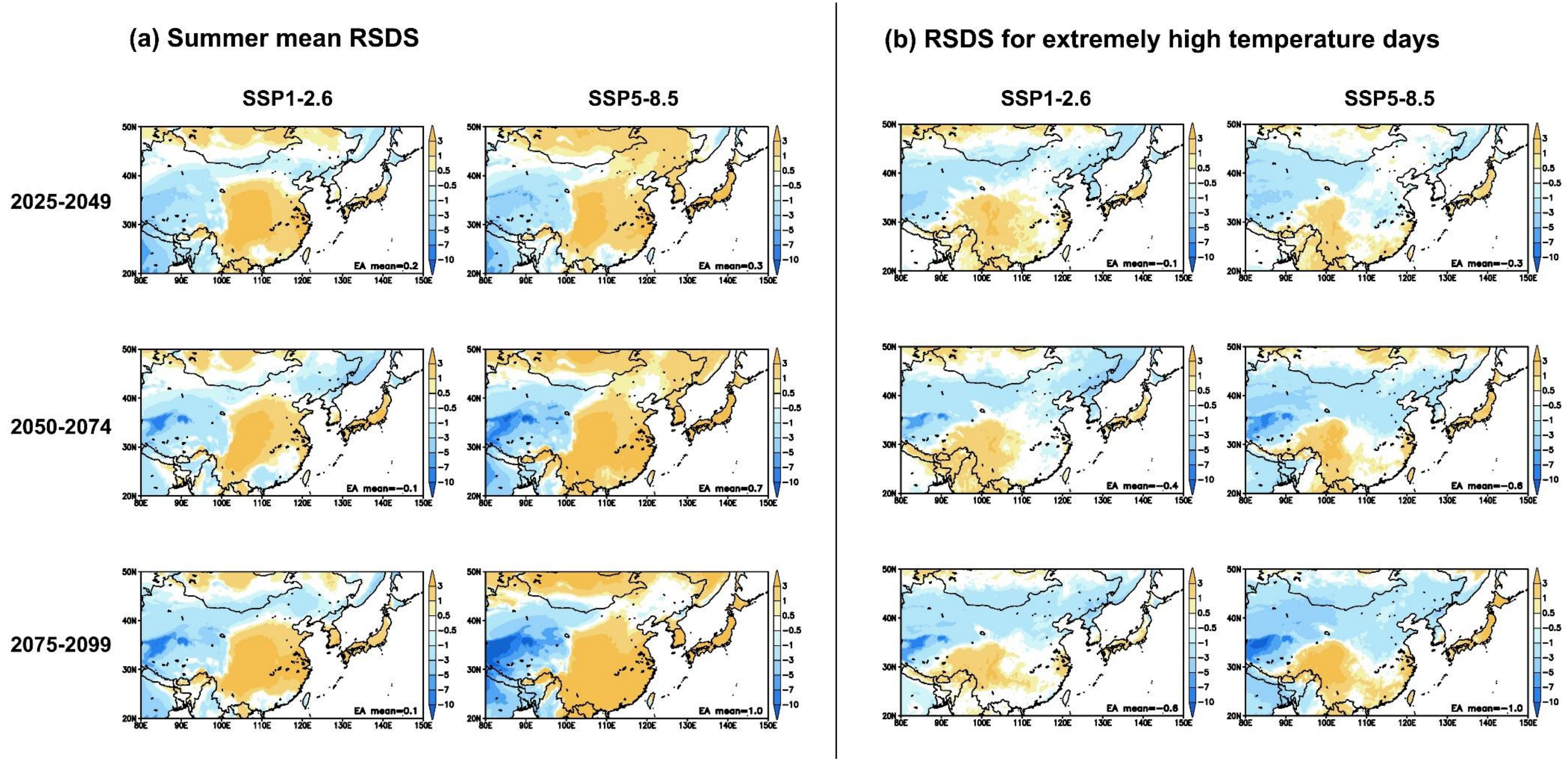

**Fig. S4.** Spatial distributions of the future changes (%) in summer mean RSDS and (b) RSDS for EHT days from three future periods and two SSP scenarios over East Asia. The area–averaged values are provided in the bottom right corners.